# Multiphasic interactions between nucleotides and target proteins

Per Nissen

Norwegian University of Life Sciences
Department of Ecology and Natural Resource Management
P. O. Box 5003, NO-1432 Ås, Norway

per.nissen@nmbu.no

2016

## Abstract

The nucleotides guanosine tetraphosphate (ppGpp) and guanosine pentaphosphate (pppGpp) bind to target proteins to promote bacterial survival (Corrigan et al. 2016). Thus, the binding of the nucleotides to RsgA, a GTPase, inhibits the hydrolysis of GTP. The dose response, taken to be curvilinear with respect to the logarithm of the inhibitor concentration, is instead much better (P<0.001 when the 6 experiments are combined) represented as multiphasic, with high to exceedingly high absolute r values for the straight lines, and with transitions in the form of non-contiguities (jumps). Profiles for the binding of radiolabeled nucleotides to HprT and Gmk, GTP synthesis enzymes, were, similarly, taken to be curvilinear with respect to the logarithm of the protein concentration. However, the profiles are again much better represented as multiphasic than as curvilinear (the P values range from 0.047 to <0.001 for each of the 8 experiments for binding of ppGpp and pppGpp to HprT). The binding of GTP to HprT and the binding of the three nucleotides to Gmk are also poorly represented by curvilinear profiles, but well represented by multiphasic profiles (straight and, in part, parallel lines).

## Introduction

In addition to multiphasic profiles for ion uptake in plants (Nissen 1971, 1974, 1991, 1996), such profiles have been recently (Nissen 2015a,b, 2016a,b,c) reported for many other processes and phenomena. In the present paper, data for the interaction between nucleotides and target proteins in Gram-positive bacteria will be reanalyzed to statistically compare the fits to curvilinear profiles with the fits to multiphasic profiles.

# Reanalysis

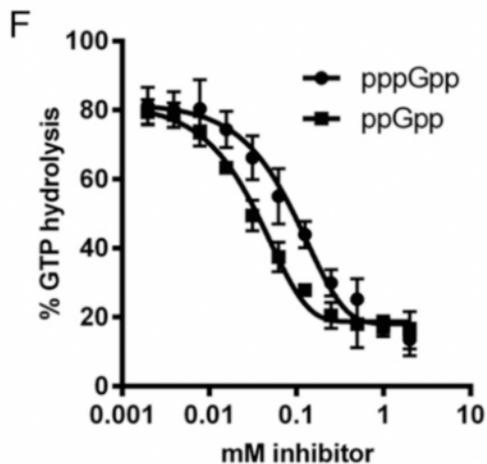

Panels A-E and their legends not shown.

Original data kindly provided by Rebecca M. Corrigan.

**Fig. 1.** Authors' Fig. 2F. Quantification of the GTPase activity of RsgA in the presence of (p)ppGpp. See also original legend.

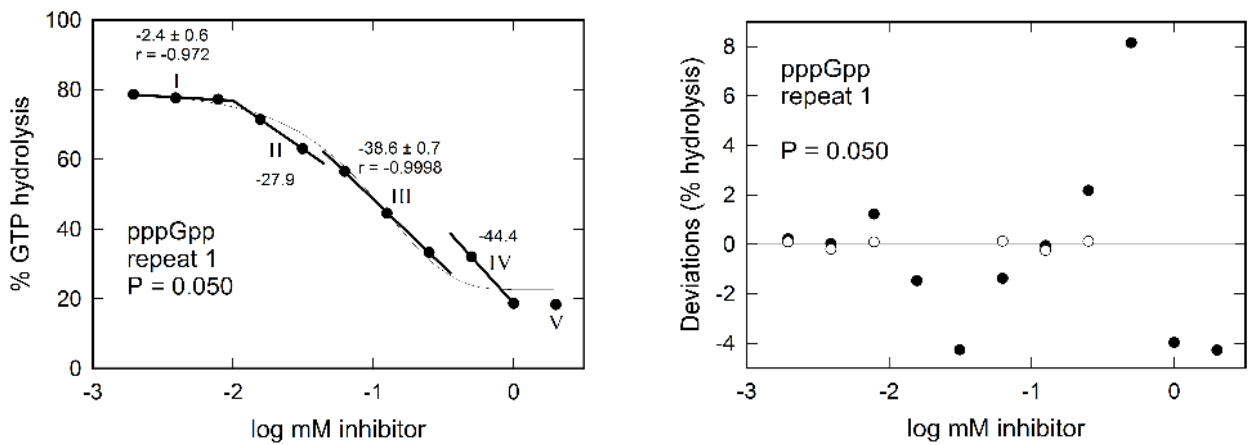

**Fig. 2** (above left). Pentaphasic profile. Transitions at -2.00, between -1.51 and -1.20 (jump), between -0.60 and -0.30 (jump), and between 0.00 and 0.30. The absolute r value for line I is quite low (for lines with shallow slopes, tiny errors can have large effects on the r values). High absolute r value for line III. Lines III and IV are about parallel.
**Fig. 3** (above right). Plot of deviates for the data in Fig. 2.

In addition to the r values, slopes ± SE (or only slopes) have been indicated. The probability that the better fit to the multiphasic profile is due to chance is also given (from Fig. 3, by the Mann-Whitney rank sum test).

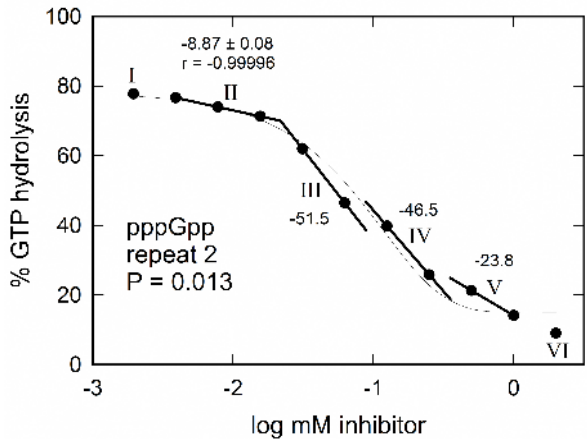 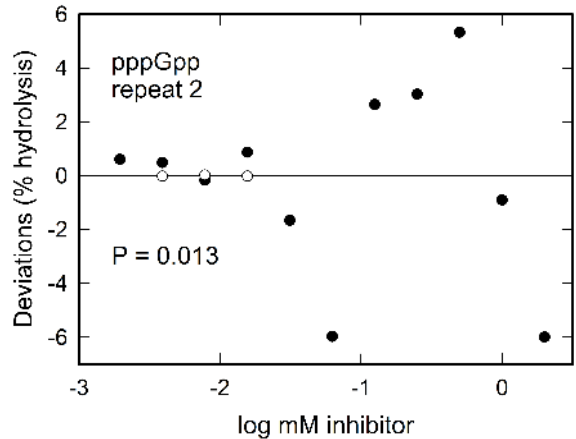

**Fig. 4** (above left). Hexaphasic profile. Transitions between -2.71 and -2.41, at -1.66, between -1.20 and -0.90 (jump), between -0.60 and -0.30 (jump), and between 0.00 and 0.30 (single lines in the range of phases I and II, and phases V and VI will be imprecise, with r values of -0.985 and -0.975, respectively). Very high absolute r value for line II. Lines III and IV are about parallel.

**Fig. 5** (above right). Plot of deviates for the data in Fig. 4.

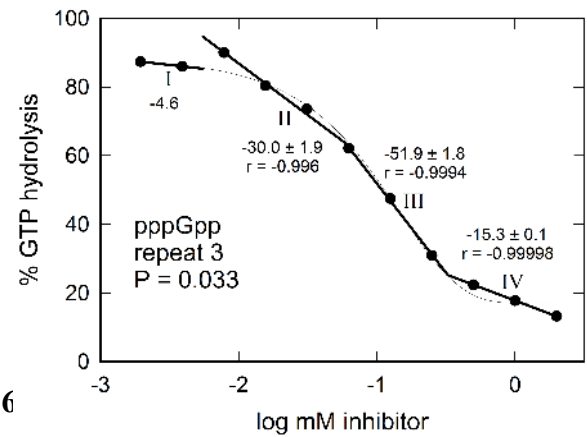 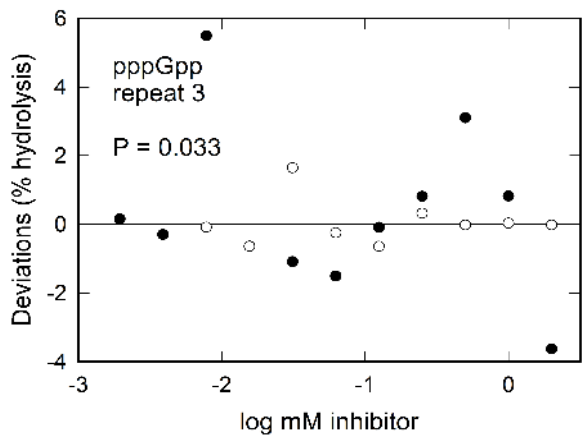

**Fig. 6** (above left). Tetraphasic profile. Transitions between -2.41 and -2.11 (jump), and at -1.20 and -0.49. Quite high absolute r value for line II, high to very high values for lines III and IV.

**Fig. 7** (above right). Plot of deviates for the data in Fig. 6.

The three profiles for pppGpp differ in the number of phases (4-6). There are parallel (and adjacent) lines in repeats 1 and 2, but not in repeat 3. There are two 3-point lines in repeat 1, one in repeat 2, and two 3-point lines and one 4-point line in repeat 3.

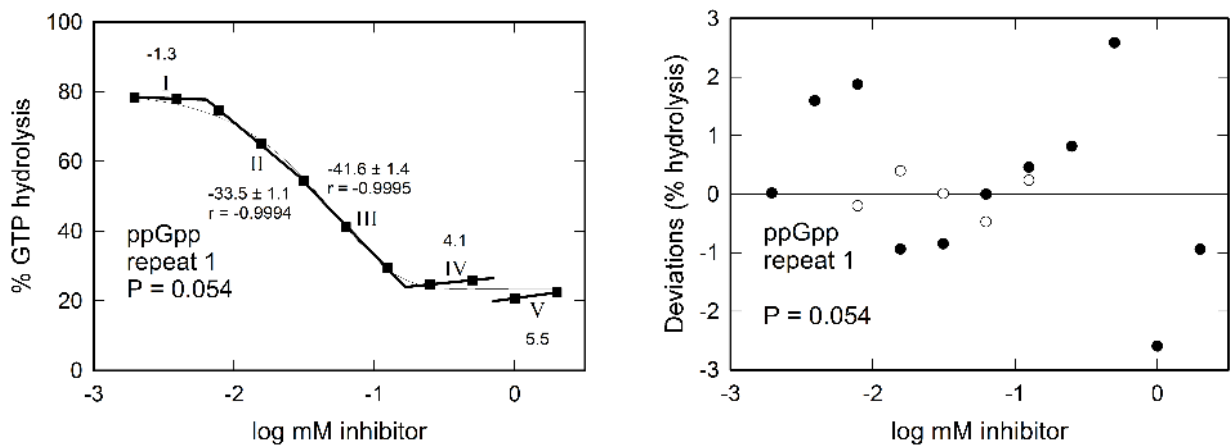

**Fig. 8** (above left). Pentaphasic profile. Transitions at -2.20, -1.51 and -0.78, and between -0.30 and 0.00 (jump). High absolute r values for lines II and III. Lines IV and V are parallel and have positive slopes.
**Fig. 9** (above right). Plot of deviates for the data in Fig. 8.

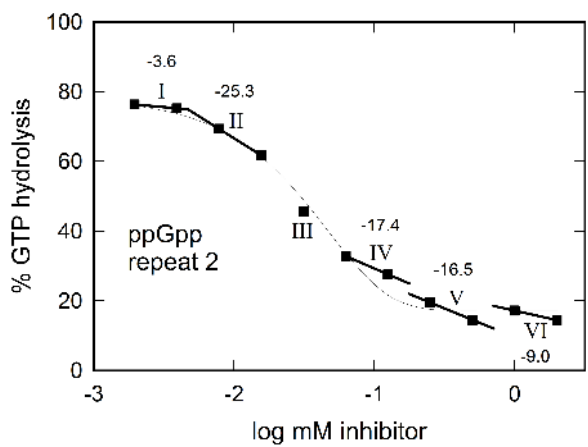

**Fig. 10.** Hexaphasic profile. Transitions at -2.33, between -1.81 and -1.51, between -1.51 and -1.20, between -0.90 and -0.60 (jump), and between -0.30 and 0.00 (jump). The data are insufficiently detailed in the range of phase III for resolution of the line. Lines IV and V are parallel. There are no lines with three or more points in the multiphasic profile, so the fits cannot be compared.

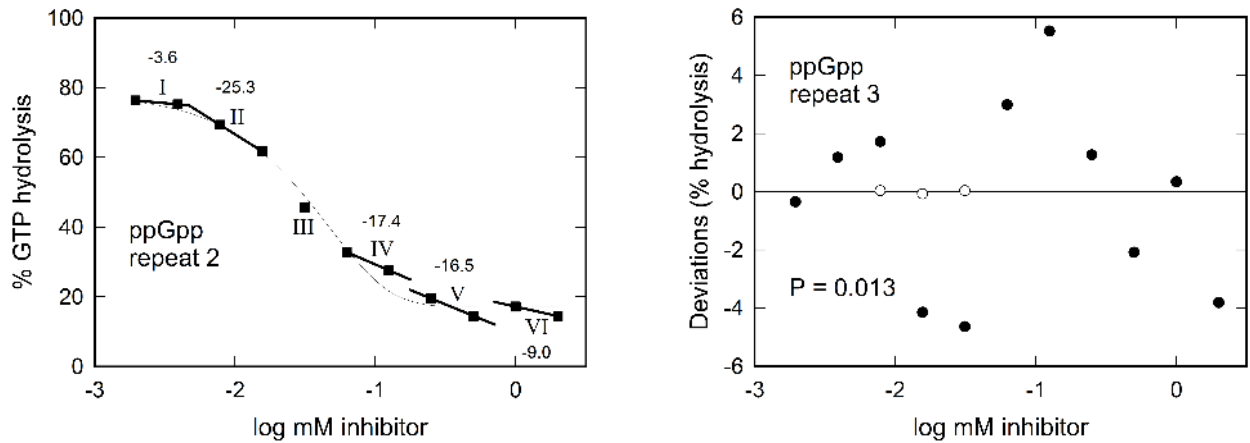

**Fig. 11** (above left). Pentaphasic profile. Transitions at -2.18, between -1.51 and -1.20 (jump), at -0.69, and between -0.30 and 0.00 (jump). Exceedingly high absolute r value for line II. Lines IV and V are parallel.
**Fig. 12** (above right). Plot of deviates for the data in Fig. 11.

As also for pppGpp, the three profiles for ppGpp differ in the number of phases (5 or 6) and in the number of 3-point lines (0-2). However, there is a set of adjacent and parallel lines in each of the profiles.

In summary, the data in Fig. 2F are well represented by multiphasic profiles. The transitions between adjacent and parallel lines are necessarily in the form of noncontiguities (jumps), and the data should not be represented by curvilinear profiles. The finding of high to exceedingly high absolute r values for the straight lines also shows that the profiles are multiphasic rather than curvilinear. P<0.001 that the better fit to multiphasic profiles is due to chance (by Fisher's (1954) method for combining independent probabilities).

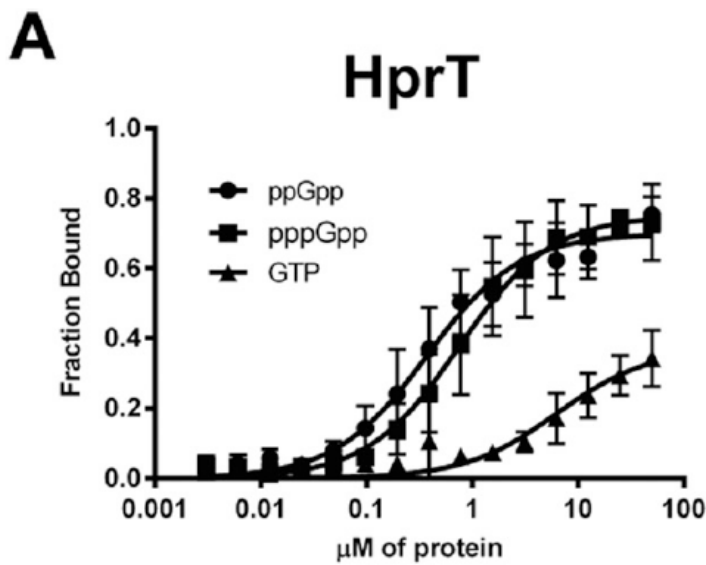

**Fig. 13.** Authors' Fig. S2A. Binding curves for radiolabeled ppGpp, pppGpp, and GTP with purified HprT$_{SA}$. For clarity, the points above are also shown in the plots below.

## Plots for HprT, ppGpp

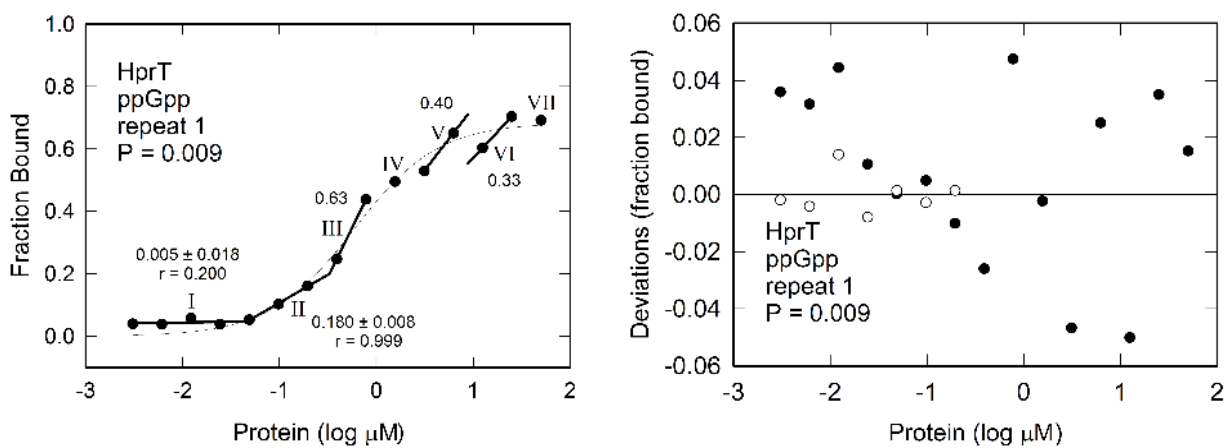

**Fig. 14** (above left). Heptaphasic profile. Transitions at -1.33 and -0.48, between -0.11 and 0.19, between 0.19 and 0.49, between 0.80 and 1.10 (jump), and between 1.40 and 1.70. The data are insufficiently detailed in the range of phase IV for resolution of the line. Line I is horizontal, line II has a high r value, and lines V and VI are about parallel. The pattern for phases IV-VII is identical to the pattern in the same range for repeat 1 for pppGpp (Fig. 22).

**Fig. 15** (above right). Plot of deviates for the data in Fig. 14.

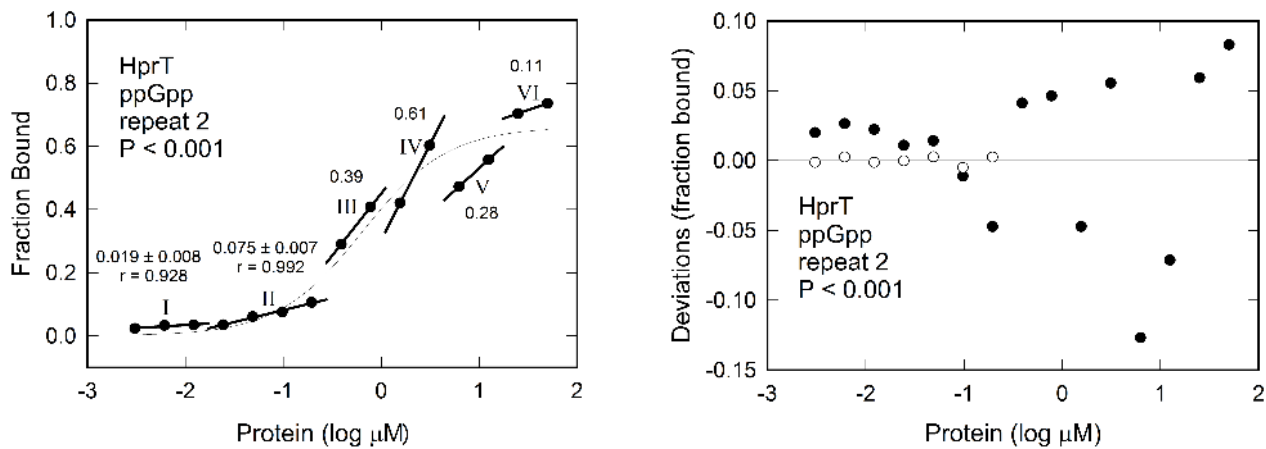

**Fig. 16** (above left). Hexaphasic profile. Transitions between -1.91 and -1.61 (jump), between -0.71 and -0.41 (jump), between -0.11 and 0.19 (jump), between 0.49 and 0.80 (jump), and between 1.10 and 1.40 (jump). Low r value for line I, but shallow slope. Quite high r value for line II.

**Fig. 17** (above right). Plot of deviates for the data in Fig. 16.

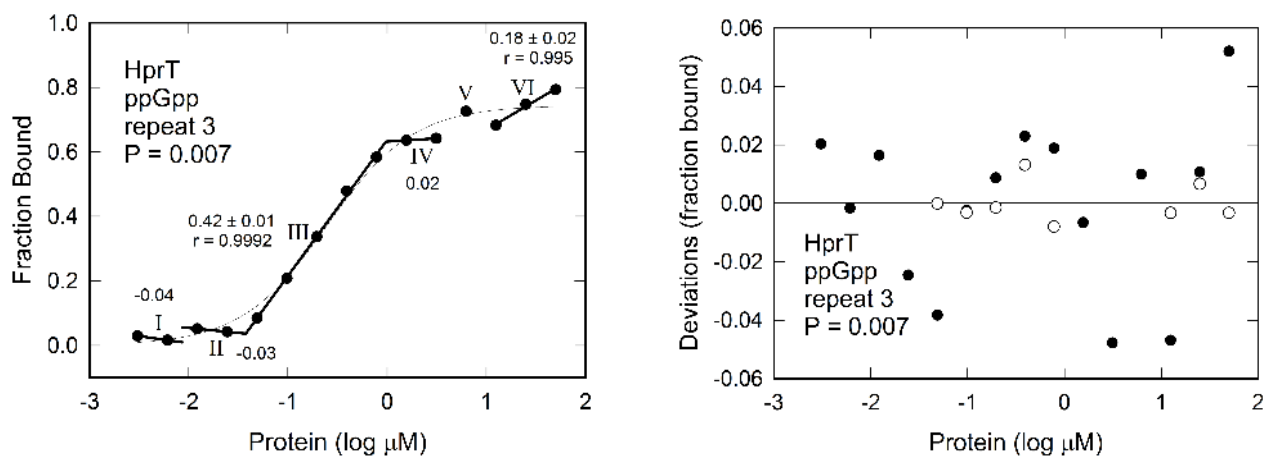

**Fig. 18** (above left). Hexaphasic profile. Transitions between -2.21 and -1.91 (jump), at -1.43 and -0.01, between 0.49 and 0.80, and between 0.80 and 1.10. The data are insufficiently detailed in the range of phase V for resolution of the line. High r values for the 5-point line III and for line VI. Lines I and II are parallel and have slightly negative slopes.

**Fig. 19** (above right). Plot of deviates for the data in Fig. 18.

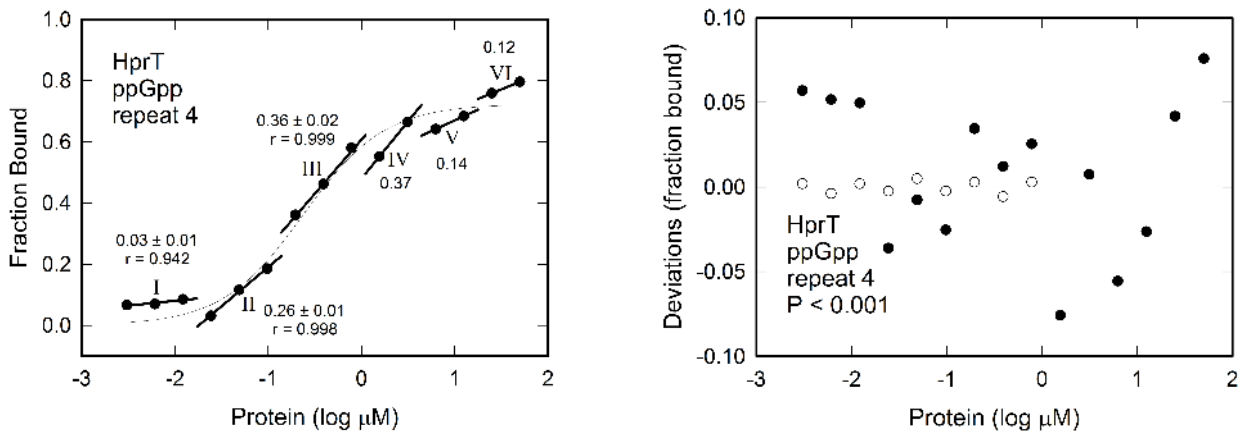

**Fig. 20** (above left). Hexaphasic profile. Transitions between -1.91 and -1.61 (jump), between -1.01 and -0.71 (jump), between -0.11 and 0.19 (jump), between 0.49 and 0.80 (jump), and between 1.10 and 1.40 (jump). Low r value for line I, but shallow slope. High r values for lines II and III. Lines III and IV are parallel, as are lines V and VI.

**Fig. 21** (above right). Plot of deviates for the data in Fig. 20.

Characteristics of multiphasic profiles for HprT, ppGpp:
Heptaphasic profile for repeat 1, hexaphasic profiles for repeats 2-4. One set of adjacent and parallel lines for repeats 1 and 3, two sets for repeat 4. Lines I and II in repeat 3 have negative slopes.

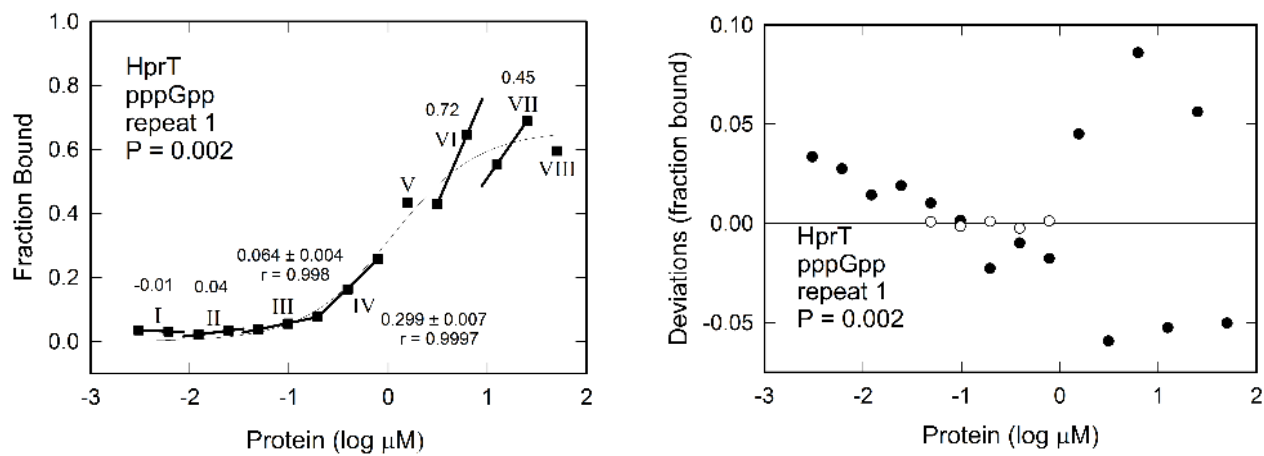

## Plots for HprT, pppGpp

**Fig. 22** (above left). Octaphasic profile. Transitions between -2.21 and -1.91 (jump), between -1.61 and -1.31 (jump), at -0.71, between -0.11 and 0.19, between 0.19 and 0.49, between 0.80 and 1.10 (jump), and between 1.40 and 1.70. The data are insufficiently detailed in the range of phase V for the line to be resolved. A single line in the range of phases I-III will have a low r value (0.790). High r values for lines III and IV. Line I has a slightly negative slope.

**Fig. 23** (above right). Plot of deviates for the data in Fig. 22.

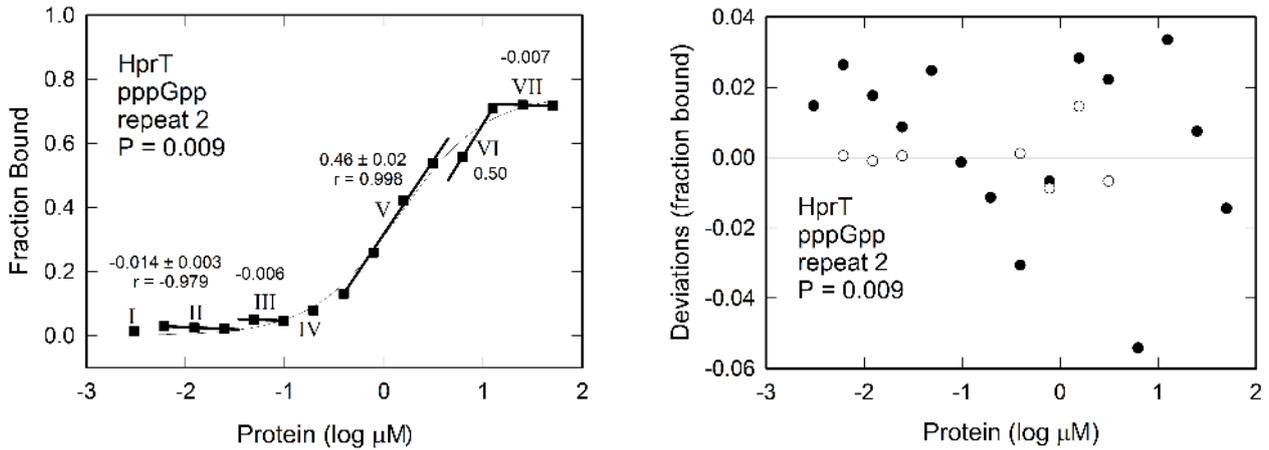

**Fig. 24** (above left). Heptaphasic profile. Transitions between -2.52 and -2.21, between -1.61 and -1.31 (jump), between -1.01 and -0.71, between -0.71 and -0.41, between 0.49 and 0.80 (jump), and at 1.12. The data are insufficiently detailed in the range of phase IV for the line to be resolved. A single line in the range of phases I-IV will have a low r value (0.869). The 4-point line V has a high r value. Line VII is horizontal. Lines II and III are parallel and have slightly negative slopes. Lines V and VI are also about parallel.
**Fig. 25** (above right). Plot of deviates for the data in Fig. 24.

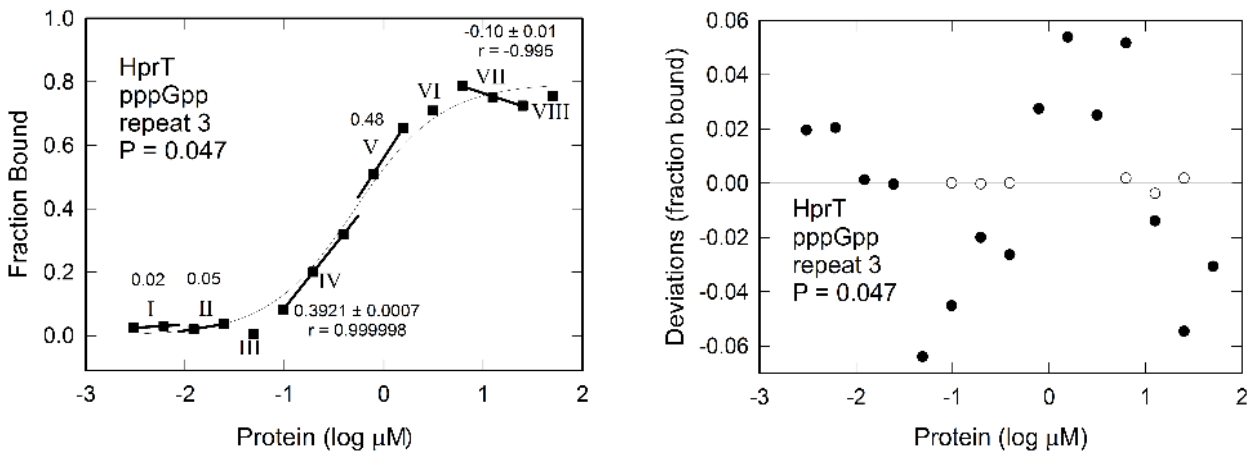

**Fig. 26** (above left). Octaphasic profile. Transitions between -2.21 and -1.91 (jump), between -1.61 and -1.31, between -1.31 and -1.01, between -0.41 and -0.11 (jump), between 0.19 and 0.49, between 0.49 and 0.80, and between 1.40 and 1.70. The data are insufficiently detailed in the range of phases III and VI for the lines to be resolved. A single line in the range of phases I-III will have r = -0.430. Exceedingly high r value for line IV. Line VII has a markedly negative slope and a quite high absolute r value.
**Fig. 27** (above right). Plot of deviates for the data in Fig. 26.

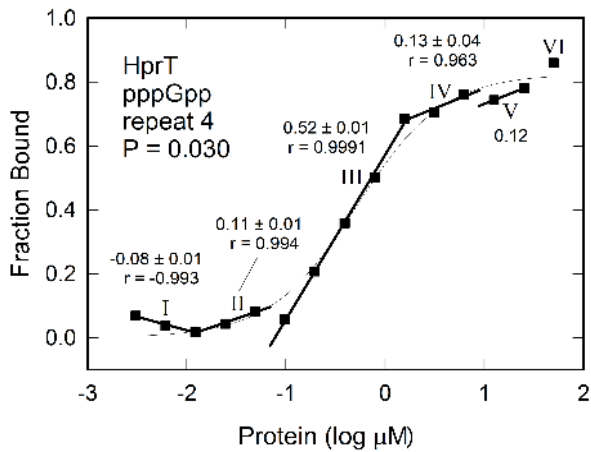 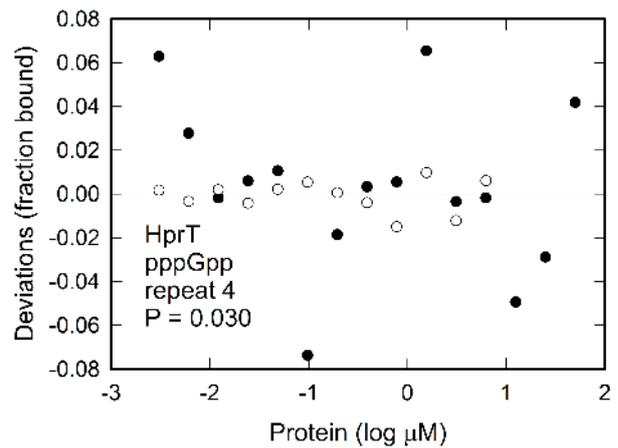

**Fig. 28** (above left). Hexaphasic profile. Transitions at -1.91, between -1.31 and -1.01 (jump), at 0.19, between 0.80 and 1.10 (jump), and between 1.40 and 1.70. Line I has a markedly negative slope and a quite high absolute r value. The r values of line II and the 5-point line III are quite high and high, respectively. The r value of line IV is quite low, but the 3-point line is parallel with line V.

**Fig. 29** (above right). Plot of deviates for the data in Fig. 28.

Characteristics of multiphasic profiles for HprT, pppGpp:
Octaphasic profiles for repeats 1 and 3, heptaphasic profile for repeat 2, hexaphasic profile for repeat 4. Two sets of adjacent and parallel lines for repeat 2, one set for repeat 4. Line I in repeat 1, lines I and VII in repeat 2, line VII in repeat 3, and line I in repeat 4 have negative slopes.

## Plots for HprT, GTP

The fits to the curvilinear profiles are clearly much poorer than the fits to the multiphasic profiles, and P values for the significance of this difference have not been calculated.

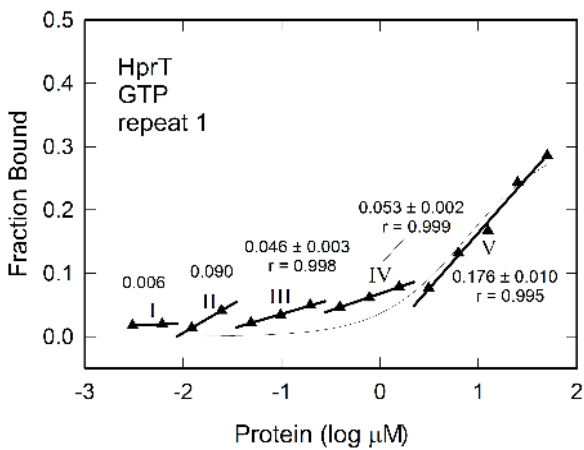 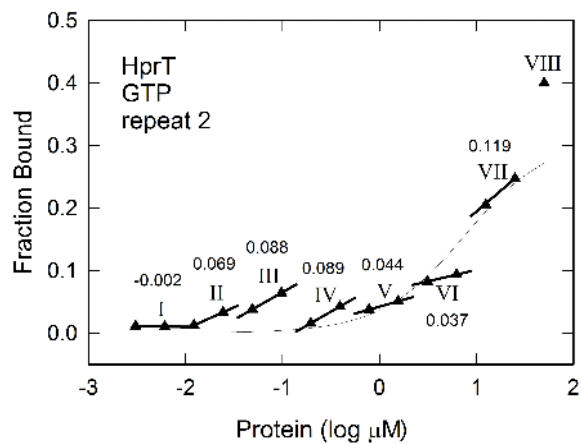

**Fig. 30** (above left). Pentaphasic profile. Transitions between -2.21 and -1.91 (jump), between -1.61 and -1.31 (jump), between -0.71 and -0.41 (jump), and between 0.19 and 0.49 (jump). High r values for lines III, IV and the 5-point line V. Lines III and IV are parallel.

**Fig. 31** (above right). Octaphasic profile. Transitions at -1.95, between -1.61 and -1.31 (jump), between -1.01 and -0.71 (jump), between -0.41 and -0.11 (jump), between 0.19 and 0.49 (jump), between 0.80 and 1.10 (jump), and between 1.40 and 1.70. Lines III and IV are parallel, as are lines V and VI.

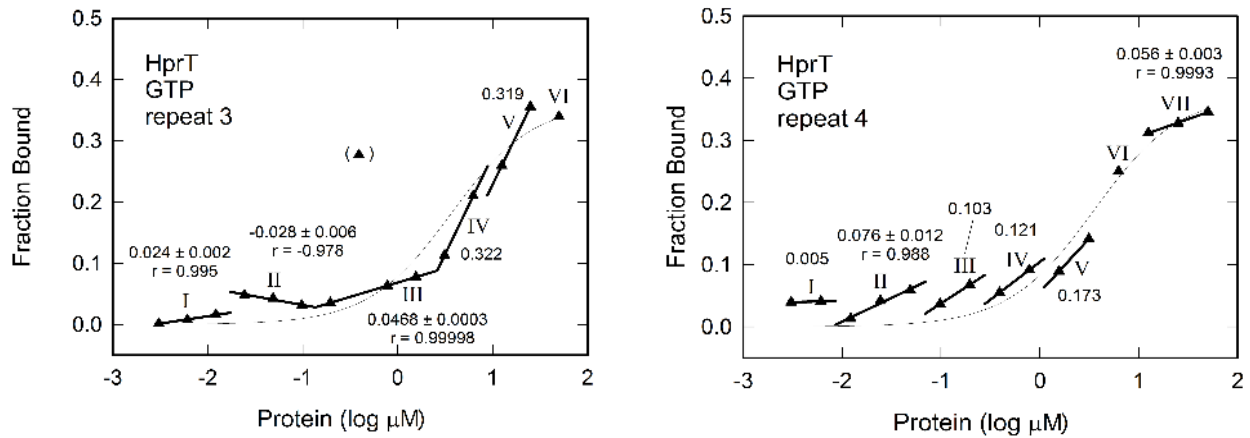

**Fig. 32** (above left). Hexaphasic profile. Transitions between -1.91 and -1.61 (jump), at -0.87 and 0.42, between 0.80 and 1.10 (jump), and between 1.40 and 1.70. High and very high r values for lines I and III, quite low absolute r value for line II which has a markedly negative slope. Lines IV and V are parallel. The value for the point at -0.41 is probably in error (see the very good fit to line III) and has been omitted from the calculations.

**Fig. 33** (above right). Heptaphasic profile. Transitions between -2.21 and -1.91 (jump), between -1.31 and -1.01 (jump), between -0.71 and -0.41 (jump), between -0.11 and 0.19 (jump), between 0.49 and 0.80, and between 0.80 and 1.10. The data are insufficiently detailed in the range of phase VI for resolution of the line. High and very high r values for lines II and VII. Lines III and IV are parallel.

Characteristics of multiphasic profiles for HprT, GTP:
The profiles are octaphasic, heptaphasic, hexaphasic and pentaphasic for repeat 2, 4, 3 and 1, respectively. Two sets of adjacent and parallel lines for repeat 2, one set for each of the other repeats. Negative slope for line II in repeat 3.

# Conclusion

From a comparison of fits it is clear, at high levels of significance, that the present data cannot be acceptably represented by curvilinear profiles. They are, instead, very well represented by multiphasic profiles, i.e. by profiles consisting of a series of straight lines separated by discontinuous transitions.

**Acknowledgment** – I am very grateful to Bob Eisenberg for his continued interest and encouragement.